\documentclass[review]{elsarticle}

\usepackage{epstopdf}
\journal{Journal of \LaTeX\ Templates}

\begin{document}

\begin{frontmatter}

\title{Theoretical design of a strain$-$controlled nanoporous CN membrane for helium separation}

%% Group authors per affiliation:
\author{Yong-Chao Rao, Zhao-Qin Chu, Xiao Gu}
\author{ Xiang-Mei Duan$^{\ast}$}
\address{Department of Physics, Faculty of Science, Ningbo University, Ningbo-315211, P.R. China}
\cortext[mycorrespondingauthor]{Corresponding author}
\ead{duanxiangmei@nbu.edu.cn}

\begin{abstract}
Designing an efficient membrane for He purification is quite crucial in scientific and industrial applications. Ultrathin membranes with intrinsic pores are highly desirable for gas purification because of their controllable aperture and homogeneous hole distribution. Based on the first$-$principles density function theory and molecular dynamics simulations, we demonstrate that the compressively strained graphitic carbon nitride (CN) can effectively purify He from Ne and Ar. Under a $-6$\% strain, the CN monolayer with a suitable pore size presents an easily surmountable barrier for He ($0.11$~eV) but formidable for Ne ($0.51$~eV) and Ar ($2.45$~eV) passing through the membrane, and it exhibits exceptionally high selectivity of $5.17\times10^6$ for He/Ne and $1.89\times10^{39}$ for He/Ar, as well as excellent He permeance of $1.94\times10^7$ GPU at room temperature, superior to those of porous graphene and C$_{\rm 2}$N membrane. Our results confirm that strain$-$tuned CN membrane could be potentially utilized for He separating from other noble gases.
\end{abstract}

\begin{keyword}
\texttt CN membrane; Helium separation; Selectivity and permeance; First$-$principles calculations; Molecular dynamics simulations
\end{keyword}

\end{frontmatter}

\section{Introduction}
With various properties, such as lower density than air, incombustibility and low index
of refraction, the lightest noble gas, helium, has been extensively applied in the fields of
semiconductors, airships and solar telescopes.\cite{Kaplan2007Helium, Cho2009Physics} However, He immediately drifts
up during the extraction from the natural gas, resulting in its irreversibly losing on
earth. And the traditional physical methods of producing He, like cryogenic distillation
or pressure$-$swing adsorption of natural gas, is hard to obtain high enough concentration
to meet the commercial utilization of He.\cite{Das2008Purification} Due to the low cost and easy operation,
membrane separation technology has been increasingly used in gas separation and
sewage purification.\cite{Pandey2001Membranes} The gas permeability are quite sensitive to the thickness of
membrane materials.\cite{D2014Helium, Kosinov2014Influence, Cao2012Helium} Therefore, the ideal membrane should be as thin as
possible to receive maximum
flux, mechanically robust to prevent cracking, and with
well$-$defined pore sizes to increase selectivity. It is highly desirable to design the efficient
He separation membrane having atomic thickness and sub$-$nanometer pores.

The single atomic thickness makes two$-$dimensional (2D) materials ideal for gas
separation compared to conventional membranes.\cite{Zhu2015C2N, Jiao2015H2} For gas purification, the hole size
of the material plays a dominated role in determining the selectivity and permeability.
Although the introduction of vacancy defects on the pristine graphene, silicene and germanene ecane
which are impermeable to gas molecules, could provide a suitable pore size for gas
separation.\cite{Bunch2008Impermeable, Hauser2012Nanoporous, Zhu2017Defective} However, it remains difficulties in controlling the size and
number of pores experimentally.\cite{Koenig2012Selective} Excitingly, some 2D porous films with uniform sub$-$nanopores, such as graphitic carbon nitride (C$_{\rm 3}$N$_{\rm4}$,\cite{Li2015Efficient} C$_{\rm 2}$N\cite{Zhu2015C2N}) and stanene,\cite{Cao2012Helium} can act as natural molecular sieves and have great potential for future He purification. To meet the increasing demand of He, researchers are focusing on
designing more efficient and accurate molecular sieves. Based on the characteristics that
porous membranes are sensitive to mechanical strain, the multi$-$stage gas separation
approach by strain$-$modulating pore size has attracted great interest.\cite{Gao2015Calculations, Silva2017Strained, Zhu2016Theoretical}

As one of the graphitic carbon nitride family, the experimentally prepared porous CN has fascinated considerable attention owing to its potential applications in energy storage and catalysts.\cite{Li2006Self, Chen2018A, Liang2016The} CN possesses honeycomb$-$like structure similar to graphene but with six$-$membered carbon$-$nitride ring in units [see Fig. 1 (a)], which makes it a candidate for gas separation.
As reported, the stability of CN was confirmed by calculating the phonon dispersion spectrum, and CN membrane could separate H$_2$ from mixed composition including CO, N$_{\rm 2}$ and CH$_{\rm 4}$.\cite{Ma2014Computational} To the best of our knowledge, research on the utilization of CN for efficient He purification is still lacking.

In this work, we propose that the CN sheet, under a biaxial compressive$-$strain of 6\%, is a good candidate for He separation.
The microscopic permeation processes of He, Ne and Ar are discussed in terms of minimum energy passway, energy profile, and electron density isosurface.
In addition, the diffusion and separation capacity are judged from permeance and selectivity at room temperature.

\section{Computational details}

The first$-$principles density functional theory (DFT) calculations are performed to optimize the structure of the porous CN, describe the electron density isosurfaces for noble gas molecules interacting with the porous CN monolayer, and carry out the energy barrier of noble gases permeating through CN membrane by using the VASP package.\cite{Kresse1996Efficiency} The perdew$-$Burke$-$Ernzerhof (PBE) functional\cite{Perdew1996Rationale} under the generalized gradient approximation (GGA) with van der Waals correction proposed by Grimme (DFT$-$D2)\cite{Grimme2010Semiempirical} is employed by the spin$-$unrestricted all$-$electron DFT calculations. The electron wave$-$functions are expanded by plane waves with cut$-$off energies of 500~eV, and the convergence criteria for electronic and ionic interactions during structure relaxation are set to be 10$^{\rm -4}$~eV and 0.01~eV, respectively. The Monkhorst$-$Pack meshes of $7\times7\times1$ are used in sampling the Brillouin zone for $2\times2$ supercells of CN, and a 20~{\AA} vacuum thickness is introduced to avoid interlayer interactions. For the transition state calculations, we have performed minimum energy path profiling using the climbing image nudged elastic band (CI$-$NEB) method\cite{Henkelman2000A} as implemented in the VASP transition state tools.

At the various temperatures, the He permeance of porous CN monolayer are investigated using MD simulations in the NVT ensemble, where the temperature of the system is controlled by the Andersen thermostat method. A condensed$-$phase optimized molecular potential for atomistic simulation studies (COMPASS) in the Material Studio software is used for describing the interatomic interaction.\cite{Sun1998COMPASS}

\section{Results and discussion}
\subsection{Pore size and stability of CN nano$-$sheet}

Figure 1(a) presents a top view of a fully relaxed $2\times2$ CN supercell. The C$-$C and C$-$N bond lengths are 1.51 and 1.34~\AA, respectively, and the optimized lattice parameter is 7.12~\AA, consistent with previous literature.\cite{Liang2016The} The hole size, with a diameter of 5.47~\AA, are illustrated by the dashed interior circle.
The electron density isosurface plot, shown in Fig. 1(b), indicates that the pore size of CN monolayer is characterized by the effective diameter of the inscribed circle, 3.28~\AA, which is much larger than the kinetic diameters of He (2.6~\AA) and a little bit larger than that of Ne molecules (3.2~\AA).\cite{Blankenburg2010Porous} Therefore, it offers the possibility to reduce the pore width in-between kinetic diameters of He and Ne molecules and makes CN sheet as an efficient He$-$isolating membrane. The biaxial compressive strain is defined as $\varepsilon=(\iota-\iota_{\rm 0})/\iota_{\rm 0}\times100\%$ , where $\iota$ and $\iota_{\rm 0}$ are the strained and original lengths of the monolayer along one of strain direction (the sample is compressed in two distinct directions with the same strain), respectively.

We investigate the He separation performance in the absence/presence of compressive strain on CN monolayer. Under compressive strain, the structural stability of CN is a key issue.
As presented in Fig. 1(c), CN can sustain a biaxial compressive strain up to $-$24\%, which means that the structure of the membrane is robust under a small compression.
Then, we figure out the cohesive energies of CN membrane under different strains. The cohesive energy is expressed by:\cite{Perim2014Inorganic}
\begin{equation}
E_{\rm {coh}}=(n_{\rm C}E_{\rm C}+n_{\rm N}E_{\rm N}-E_{\rm CN})/(n_{\rm C}+n_{\rm N})
\end{equation}
Where $E_{\rm C}$, $E_{\rm N}$ and $E_{\rm CN}$ are the energy of single C atom, single N atom, total CN membrane, respectively; and $n_{\rm C}$, $n_{\rm N}$ are the total number of C and N atoms. Fig. 1(c) shows that $E_{\rm coh}$ decreases with the increasing strain, while $E_{\rm coh}$ under 6\% compressive strain is 6.05~eV per atom, which is still much lower than that of silicene (3.17~eV per atom).\cite{Li2014Be} And the strain energy of the system ($E_{\rm s}$, the energy difference between strained and equilibrium states), which actually equals to $E_{\rm {s}}=-[E_{\rm coh}(\varepsilon_i)-E_{\rm coh}(\varepsilon_0)]\times(n_{\rm C}+n_{\rm N})$, increases with strain monotonically until the strain reaches the critical value of $-$24\%.
Note that silicene can be successfully used as gas separation membrane,\cite{Hu2013Helium} therefore CN under moderate mechanical strain is stable enough for He separation.

\subsection{Diffusion energy barrier of the noble molecule}

Before exploring the transition state and energy barrier, we first optimize the energetically most stable state (SS) of gas molecules on the surface of CN monolayer.
The atomic structures of He atom adsorbed on CN surface with and without strain are presented in Fig. 2.
The interaction energy between a noble gas molecule and CN membrane is defined by:
\begin{equation}
E_{\rm {int}}=E_{\rm {gas/CN}}-E_{\rm gas}-E_{\rm CN}
\end{equation}
Where $E_{\rm {gas/CN}}$, $E_{\rm gas}$ and $E_{\rm CN}$ represent the total enegy of gas/CN system, isolated gas molecule and pure CN monolayer, respectively.
The interaction energy and adsorption height of the SS under different compressive strains are summarized in Table 1. The interaction energy and adsorption height are in the range of $-$6.9 to $-$25.5~meV and 1.05 to 2.61~\AA, respectively, indicating the gas molecules are all physically adsorbed on the CN monolayer via weak van der Waals interaction.
The magnitude of adsorption energy is comparable to those of the noble gases adsorbed on C$_{\rm 3}$N$_{\rm4}$ ($-$17 to $-$99~meV) and C$_{\rm 2}$N sheet ($-$60 to $-$90 meV).
The general trend is that as the compressive strain increases, the reduction in pore size would enhance the repulsion interaction between the gas molecules and the CN membrane, leading to an increase in molecular adsorption height.
Specifically, Ar has the highest adsorption height, which is due to the relatively large kinetic diameter.

When the gas migrates from the most stable adsorption site to the CN film, due to the symmetry, the transition state (TS) should be which the gas molecules are at the center of the cavity and in the same plane as the film. The TS state is confirmed using the NEB approach.
The energy barrier (E$_{\rm b}$) is defined as:
\begin{equation}
E_{\rm {b}}=E_{\rm TS}-E_{\rm SS}
\end{equation}
Where $E_{\rm TS}$ and $E_{\rm SS}$ represents the energy between gas molecules and CN membrane at the TS and SS, respectively. The energy profiles and barriers of He, Ne and Ar passing through the pore under different compressive strain are shown in Fig. 3. The barrier for He, Ne and Ar passing through the pristine CN is 0.03, 0.18 and 1.13~eV, respectively. Clearly, the relatively low penetration barrier for both He and Ne make it impracticable to purify He via unstrained CN membrane. Fortunately, the energy barrier increases as the compressive strain is exerted. In particular, when the strain is reached to $-$6\%, the barrier for He, Ne and Ar is further increased to 0.11, 0.51 and 2.45~eV, respectively. The unexpectedly lower penetration barrier for He is within the threshold barrier for gas penetration (0.50~eV),\cite{Schrier2010Helium} while the values for the other two noble gases are lager than 0.50~eV, suggesting that the He separation performance of CN can be significantly improved by inducing a 6\% compressive strain.

To deeply understand of the change of energy barrier when the molecules pass strain$-$modified CN film, we show the electron density isosurface of noble gases interacting with CN at TS under $-$6\% strain. Intuitively, large electron density overlap would induce strong repulsion which hinder the movement of gas molecules through the pore. From Fig. 4(a), it can be seen that there is no overlap between the electron density of He and the CN, therefore He could easily move through the membrane. However, the electron density overlap increases for Ne and Ar, as presented in Fig. 4(b) and (c), the Ar molecule and the porous CN sheet completely overlap in electron density, so the Ar gas molecules cannot pass through the film.
The effective diameter of strained CN [shown in the inscribed circle of Fig. 4(a)] is now 2.99~\AA, which is between the kinetic diameter of He and Ne. Hence the compressed membrane only allows He to penetrate.

The He separation efficiency of the porous CN membrane can be quantitatively examined through the Arrhenius$-$equation.\cite{Jiang2009Porous}
The selectivity (S) for He over other two noble gas molecules is estimated by:
\begin{equation}
S_{\rm {He/gas}}=\frac{{\gamma}_{\rm He}}{{\gamma}_{\rm gas}}=\frac{A_{\rm He}exp(-E_{\rm He}/RT)}{A_{\rm gas}exp(-E_{\rm gas}/RT)}
\end{equation}
Where $\gamma$ is the diffusion rate, the diffusion prefactor, $A$, is set to $10^{11}~s^{-1}$,\cite{Blankenburg2010Porous} $R$ and $T$ are the Boltzmann constant and the absolute temperature, respectively. $E$ is the diffusion energy barrier. The temperature$-$dependent diffusion rates of noble gas molecules and He selectivity relative to Ne and Ar are depicted in Fig. 5(a) and (b), respectively.
Generally, with the temperature increasing, the diffusion rate of the molecules increases, while the He selectivity of CN membrane relative to Ne and Ar decreases. The main reason is that, when the temperature rises, the kinetic energy of the inert gas molecule is greatly increased, make it possible to overcome the energy barriers and pass the CN film. Significantly, at a certain temperature, the enhanced compressive strain reduces down the He diffusion rate [Fig. 5(a)] but greatly increases the selectivity for He/Ne and He/Ar [Fig. 5(b)], which demonstrates the appropriate compressive strain can effectively accelerate the separation of He from Ne and Ar. As summarized in Table 2, at room temperature, the selectivity for He/Ne and He/Ar are about $5.17\times10^6$ and $1.89\times10^{39}$ by subjecting to a $-$6\% strain, respectively. The values are much superior than that of other theoretical results and surely acceptable for the industrial application.\cite{Zhu2015C2N,Hu2013Helium,Zhu2006Permeance}

\subsection{He separation by MD simulations}

Permeability is another critical criterion to characterize the separation performance of CN membrane. Based on MD simulations, we investigate the gas flow to estimate the permeability of CN membrane quantitatively at room temperature, which is defined as:\cite{Du2011Separation}
\begin{equation}
F=\frac{\nu}{S\times{t}\times{\Delta{P}}}
\end{equation}
Where $\nu$ and $S$ represent the moles of gas molecules in the permeate side and the area of CN membrane, $t$ is the time duration, and the pressure drop ($\Delta{P}$) is set to 1~{bar} across the pore.

As shown in the Fig. 6, the $4\times4$ supercell with a height of 90~{\AA} is divided into three parts of the same capacity. Periodic boundary conditions along the lateral directions are imposed to mimic the infinite film. At the beginning of each simulation, the gas reservoir, consisting of 40 He molecules, 40 Ne molecules and 40 Ar molecules, is initially located in the middle part of the supercell, that is, between the two monolayers of CN. The van der Waals interactions and Ewald electrostatic interactions are applied with a cutoff distance of 9.5~\AA. Each simulation is carried out for a time period of 10~$ns$ with a time$-$step of 1~$fs$.
In the process of simulations,
due to the existing of difference in gas concentration among different spaces, the molecules are pushed cross the sheet into the vacuum region. After 10~$ns$ simulation, the final configurations under 0\%, $-$3\% and $-$6\% strain at room temperature are shown in Fig. 6. Some of Ne atoms pass through the pristine CN membrane [see Fig. 6(a)], and there are still several Ne atoms penetrate the membrane under a small strain of $-$3\% [see Fig. 6(b)], which are consistent with the low penetration barriers in Fig. 3(a) and (b). However, under a $-$6\% strain [see Fig. 6(c)], 27 He in 40 molecules pass through CN film to the vacuum space, without any Ne and Ar penetrating.
The MD simulations further confirm that the strain$-$controlled nanoporous CN membrane can be applied in the purification of He from other noble gas molecules.

The calculated He permeance of porous CN monolayer under $-$6\% strain with area of $28.48\times28.48$~{\AA}$^{\rm 2}$ together with that of the previously proposed porous monolayer are summarized in Table 3.
The strain$-$modified CN membrane exhibits a He permeability as high as $1.94\times10^7$~GPU at 300~$K$, which is almost twice that of C$_{\rm 2}$N, and 270 times greater than that of the porous graphene (PG), far much higher than the industrially acceptable gas separation value of 20.\cite{Zhu2006Permeance}

\section{Conclusions}
Combining DFT$-$D2 calculations and classical MD simulation, we predict that strained porous CN monolayer can be used for He separation with high selectivity and permeability. The pristine CN is permeable for both He and Ne because of the lower barriers than threshold value. The energy barriers of gas molecules can be effectively modulated by biaxial compressive strain to the membrane. Subjecting the CN membrane to a $-$6\% strain results in non$-$passability of Ne and exceptionally high selectivity for He/Ne (Ar) at 300~K. Meanwhile, the MD simulations further verify that the strained CN membrane exhibits a high He permeance of $1.94\times10^7$~GPU, which is superior to those of PG and C$_2$N sheets. Our results demonstrate that strain is an efficient strategy to tune the separation performance of low$-$dimensional materials and reveal that strain$-$controlled CN monolayer has great potential application in He separation.   	

\section*{Acknowledgments}

This research is supported by the Natural Science Foundation of China (grant No. 11574167), the New Century 151 Talents Project of Zhejiang Province and the KC Wong Magna Foundation in Ningbo University.

\section*{Conflict of interest}
The authors declare they have no conflict of interest

\section*{References}

\bibliography{references}

\begin{thebibliography}{10}
\expandafter\ifx\csname url\endcsname\relax
  \def\url#1{\texttt{#1}}\fi
\expandafter\ifx\csname urlprefix\endcsname\relax\def\urlprefix{URL }\fi
\expandafter\ifx\csname href\endcsname\relax
  \def\href#1#2{#2} \def\path#1{#1}\fi

\bibitem{Kaplan2007Helium}
K.~H. Kaplan, Helium shortage hampers research and industry, Phys. Today 60
  (2007) 31--32.
\newblock \href {http://dx.doi.org/10.1063/1.2754594}
  {\path{doi:10.1063/1.2754594}}.

\bibitem{Cho2009Physics}
A.~Cho, Helium$-$3 shortage could put freeze on low$-$temperature research.,
  Science 326 (2009) 778--779.
\newblock \href {http://dx.doi.org/10.1126/science.326-778}
  {\path{doi:10.1126/science.326-778}}.

\bibitem{Das2008Purification}
N.~K. Das, H.~Chaudhuri, R.~K. Bhandari, D.~Ghose, P.~Sen, B.~Sinha,
  \href{http://www.jstor.org/stable/24105328}{Purification of helium from
  natural gas by pressure swing adsorption}, Curr. Sci. 95 (2008) 1684--1687.
\newline\urlprefix\url{http://www.jstor.org/stable/24105328}

\bibitem{Pandey2001Membranes}
P.~Pandey, R.~S. Chauhan, Membranes for gas separation, Prog. Polym. Sci. 26
  (2001) 853--893.
\newblock \href {http://dx.doi.org/org/10.1016/S0079-6700(01)00009-0}
  {\path{doi:org/10.1016/S0079-6700(01)00009-0}}.

\bibitem{D2014Helium}
B.~D\'iez, P.~Cuadrado, A.~MarcosFern\'andez, P.~Pr\'adanos, A.~Tena,
  L.~Palacio, A.~E. Lozano, A.~Hern\'andez, Helium recovery by membrane gas
  separation using poly(o$-$acyloxyamide)s, Ind. Eng. Chem. Res. 53 (2014)
  12809--12818.
\newblock \href {http://dx.doi.org/10.1021/ie501649b}
  {\path{doi:10.1021/ie501649b}}.

\bibitem{Kosinov2014Influence}
N.~Kosinov, C.~Auffret, V.~G.~P. Sripathi, C.~G\"uc\"uyener, J.~Gascon,
  F.~Kapteijn, E.~J.~M. Hensen, Influence of support morphology on the
  detemplation and permeation of {ZSM}$-$5 and {SSZ}$-$13 zeolite membranes,
  Microporous Mesoporous Mater. 197 (2014) 268--277.
\newblock \href {http://dx.doi.org/10.1016/j.micromeso.2014.06.022}
  {\path{doi:10.1016/j.micromeso.2014.06.022}}.

\bibitem{Cao2012Helium}
F.~Cao, C.~Zhang, Y.~Xiao, H.~Huang, W.~Zhang, D.~Liu, C.~Zhong, Q.~Yang,
  Z.~Yang, X.~Lu, Helium recovery by a {C}u$-${BTC} metal$-$organic$-$framework
  membrane, Ind. Eng. Chem. Res. 51 (2012) 11274--11278.
\newblock \href {http://dx.doi.org/10.1021/ie301445p}
  {\path{doi:10.1021/ie301445p}}.

\bibitem{Zhu2015C2N}
L.~Zhu, Q.~Xue, X.~Li, T.~Wu, Y.~Jin, W.~Xing, {C$_{\rm 2}$N}: an excellent
  two$-$dimensional monolayer membrane for {H}e separation, J. Mater. Chem. A 3
  (2015) 21351--21356.
\newblock \href {http://dx.doi.org/10.1039/C5TA05700K}
  {\path{doi:10.1039/C5TA05700K}}.

\bibitem{Jiao2015H2}
Y.~Jiao, A.~Du, S.~C. Smith, Z.~Zhu, S.~Qiao, H$_{\rm 2}$ purification by
  functionalized graphdiyne$-$role of nitrogen doping, J. Mater. Chem. A 3
  (2015) 6767--6771.
\newblock \href {http://dx.doi.org/10.1039/C5TA01062D}
  {\path{doi:10.1039/C5TA01062D}}.

\bibitem{Bunch2008Impermeable}
J.~S. Bunch, S.~S. Verbridge, J.~S. Alden, A.~M. v.~d. Zande, J.~M. Parpia,
  H.~G. Craighead, P.~L. McEuen, Impermeable atomic membranes from graphene
  sheets, Nano Lett. 8 (2008) 2458--2462.
\newblock \href {http://dx.doi.org/10.1021/nl801457b}
  {\path{doi:10.1021/nl801457b}}.

\bibitem{Hauser2012Nanoporous}
A.~W. Hauser, P.~Schwerdtfeger, Nanoporous graphene membranes for efficient
  ${^3}${H}e/${^4}${H}e separation, J. Phys. Chem. Lett. 3~(2) (2012) 209--213.
\newblock \href {http://dx.doi.org/10.1021/jz201504k}
  {\path{doi:10.1021/jz201504k}}.

\bibitem{Zhu2017Defective}
L.~Zhu, X.~Chang, D.~He, Q.~Xue, X.~Li, Y.~Jin, H.~Zheng, C.~Ling, Defective
  germanene as a high$-$efficiency helium separation membrane: a
  first$-$principles study, Nanotech. 28 (2017) 135703.
\newblock \href {http://dx.doi.org/10.1088/1361-6528/aa5fae}
  {\path{doi:10.1088/1361-6528/aa5fae}}.

\bibitem{Koenig2012Selective}
S.~P. Koenig, L.~Wang, J.~Pellegrino, J.~S. Bunch, Selective molecular sieving
  through porous graphene, Nat. Nanotech. 7 (2012) 728--732.
\newblock \href {http://dx.doi.org/10.1038/nnano.2012.162}
  {\path{doi:10.1038/nnano.2012.162}}.

\bibitem{Li2015Efficient}
F.~Li, Y.~Qu, M.~Zhao, Efficient helium separation of graphitic carbon nitride
  membrane, Carbon 95 (2015) 51--57.
\newblock \href {http://dx.doi.org/10.1016/j.carbon.2015.08.013}
  {\path{doi:10.1016/j.carbon.2015.08.013}}.

\bibitem{Gao2015Calculations}
G.~Gao, Y.~Jiao, F.~Ma, L.~Kou, A.~Du, Calculations of helium separation via
  uniform pores of stanene$-$based membranes, Beilstein J. Nanotech. 6 (2015)
  2470--2476.
\newblock \href {http://dx.doi.org/10.3762/bjnano.6.256}
  {\path{doi:10.3762/bjnano.6.256}}.

\bibitem{Silva2017Strained}
S.~W.~D. Silva, A.~Du, W.~Senadeera, Y.~Gu, Strained graphitic carbon nitride
  for hydrogen purification, J. Membr. Sci. 528 (2017) 201--205.
\newblock \href {http://dx.doi.org/10.1016/j.memsci.2017.01.034}
  {\path{doi:10.1016/j.memsci.2017.01.034}}.

\bibitem{Zhu2016Theoretical}
L.~Zhu, Y.~Jin, Q.~Xue, X.~Li, H.~Zheng, T.~Wu, C.~Ling, Theoretical study of a
  tunable and strain$-$controlled nanoporous graphenylene membrane for
  multifunctional gas separation, J. Mater. Chem. A 4 (2016) 15015--15021.
\newblock \href {http://dx.doi.org/10.1039/C6TA04456E}
  {\path{doi:10.1039/C6TA04456E}}.

\bibitem{Li2006Self}
J.~Li, C.~Cao, J.~Hao, H.~Qiu, Y.~Xu, H.~Zhu, Self$-$assembled
  one$-$dimensional carbon nitride architectures, Diamond Relat. Mater. 15
  (2006) 1593--1600.
\newblock \href {http://dx.doi.org/10.1016/j.diamond.2006.01.013}
  {\path{doi:10.1016/j.diamond.2006.01.013}}.

\bibitem{Chen2018A}
Y.~D. Chen, S.~Yu, W.~H. Zhao, S.~F. Li, X.~M. Duan, A potential material for
  hydrogen storage: a {L}i decorated graphitic$-${CN} monolayer, Phys. Chem.
  Chem. Phys. 20 (2018) 13473--13477.
\newblock \href {http://dx.doi.org/10.1039/C8CP01145A}
  {\path{doi:10.1039/C8CP01145A}}.

\bibitem{Liang2016The}
D.~Liang, T.~Jing, Y.~Ma, J.~Hao, G.~Sun, M.~Deng, The photocatalytic
  properties of g$-${C}$_{\rm 6}${N}$_{\rm 6}$/g$-${C}$_{\rm 3}${N}$_{\rm 4}$
  heterostructure: a theoretical study, J. Phys. Chem. C 120 (2016)
  24023--24029.
\newblock \href {http://dx.doi.org/10.1021/acs.jpcc.6b08699}
  {\path{doi:10.1021/acs.jpcc.6b08699}}.

\bibitem{Ma2014Computational}
Z.~Ma, X.~Zhao, Q.~Tang, Z.~Zhen, Computational prediction of experimentally
  possible g$-${C}$_{\rm 3}${N}$_{\rm 3}$ monolayer as hydrogen purification
  membrane, Int. J. of Hydrogen Energy 39 (2014) 5037--5042.
\newblock \href {http://dx.doi.org/10.1016/j.ijhydene.2014.01.046}
  {\path{doi:10.1016/j.ijhydene.2014.01.046}}.

\bibitem{Kresse1996Efficiency}
G.~Kresse, J.~Furthm\"uller, Efficiency of ab$-$initio total energy
  calculations for metals and semiconductors using a plane$-$wave basis set,
  Comp. Mater. Sci. 6 (1996) 15--50.
\newblock \href {http://dx.doi.org/10.1016/0927-0256(96)00008-0}
  {\path{doi:10.1016/0927-0256(96)00008-0}}.

\bibitem{Perdew1996Rationale}
J.~P. Perdew, M.~Ernzerhof, K.~Burke, Rationale for mixing exact exchange with
  density functional approximations, J. Chem. Phys. 105 (1996) 9982--9985.
\newblock \href {http://dx.doi.org/10.1063/1.472933}
  {\path{doi:10.1063/1.472933}}.

\bibitem{Grimme2010Semiempirical}
S.~Grimme, Semiempirical {GGA}$-$type density functional constructed with a
  lon$-$range dispersion correction, J. Comp. Phys. 27 (2010) 1787--1799.
\newblock \href {http://dx.doi.org/10.1002/jcc.20495}
  {\path{doi:10.1002/jcc.20495}}.

\bibitem{Henkelman2000A}
G.~Henkelman, A climbing image nudged elastic band method for finding saddle
  points and minimum energy paths, J. Chem. Phys. 113 (2000) 9901--9904.
\newblock \href {http://dx.doi.org/10.1063/1.1329672}
  {\path{doi:10.1063/1.1329672}}.

\bibitem{Sun1998COMPASS}
H.~Sun, Compass: An ab initio force$-$field optimized for condensed-phase
  applications$-$overview with details on alkane and benzene compounds, J.
  Phys. Chem. B 102 (1998) 7338--7364.
\newblock \href {http://dx.doi.org/10.1021/jp980939v}
  {\path{doi:10.1021/jp980939v}}.

\bibitem{Blankenburg2010Porous}
S.~Blankenburg, M.~Bieri, R.~Fasel, K.~M\"ullen, C.~A. Pignedoli, D.~Passerone,
  Porous graphene as an atmospheric nanofilter, Small 6 (2010) 2266.
\newblock \href {http://dx.doi.org/10.1002/smll.201001126}
  {\path{doi:10.1002/smll.201001126}}.

\bibitem{Perim2014Inorganic}
E.~Perim, R.~Paupitz, P.~A.~S. Autreto, D.~S. Galvao, Inorganic graphenylene: a
  porous two$-$dimensional material with tunable band gap, J. Phys. Chem. C 118
  (2014) 23670--23674.
\newblock \href {http://dx.doi.org/10.1021/jp502119y}
  {\path{doi:10.1021/jp502119y}}.

\bibitem{Li2014Be}
Y.~Li, Y.~Liao, Z.~Chen, Be$_{\rm 2}${C} monolayer with quasi$-$planar
  hexacoordinate carbons: a global minimum structure, Angew. Chem. Int. Ed. 53
  (2014) 7248--7252.
\newblock \href {http://dx.doi.org/10.1002/anie.201403833}
  {\path{doi:10.1002/anie.201403833}}.

\bibitem{Hu2013Helium}
W.~Hu, X.~Wu, Z.~Li, J.~Yang, Helium separation via porous silicene based
  ultimate membrane, Nanoscale 5 (2013) 9062--9066.
\newblock \href {http://dx.doi.org/10.1039/C3NR02326E}
  {\path{doi:10.1039/C3NR02326E}}.

\bibitem{Schrier2010Helium}
J.~Schrier, Helium separation using porous graphene membranes, J. Phys. Chem.
  Lett. 1 (2010) 2284--2287.
\newblock \href {http://dx.doi.org/10.1021/jz100748x}
  {\path{doi:10.1021/jz100748x}}.

\bibitem{Jiang2009Porous}
D.~Jiang, V.~R. Cooper, S.~Dai, Porous graphene as the ultimate membrane for
  gas separation, Nano Lett. 9 (2009) 4019--4024.
\newblock \href {http://dx.doi.org/10.1021/nl9021946}
  {\path{doi:10.1021/nl9021946}}.

\bibitem{Zhu2006Permeance}
Z.~Zhu, Permeance should be used to characterize the productivity of a
  polymeric gas separation membrane, J. Membr. Sci. 281 (2006) 754--756.
\newblock \href {http://dx.doi.org/10.1016/j.memsci.2006.04.040}
  {\path{doi:10.1016/j.memsci.2006.04.040}}.

\bibitem{Du2011Separation}
H.~Du, J.~Li, J.~Zhang, G.~Su, X.~Li, Y.~Zhao, Separation of hydrogen and
  nitrogen gases with porous graphene membrane, J. Phys. Chem. C 115 (2011)
  23261--23266.
\newblock \href {http://dx.doi.org/10.1021/jp206258u}
  {\path{doi:10.1021/jp206258u}}.

\bibitem{Brockway2013Noble}
A.~M. Brockway, J.~Schrier, Noble gas separation using {PG$-$ES$X$} ({$X$} = 1,
  2, 3) nanoporous two$-$dimensional polymers, J. Phys. Chem. C 117 (2013)
  393--402.
\newblock \href {http://dx.doi.org/10.1021/jp3101865}
  {\path{doi:10.1021/jp3101865}}.

\end{thebibliography}
\newpage
\begin{figure}[h]
  \centering
  \includegraphics[width=9cm]{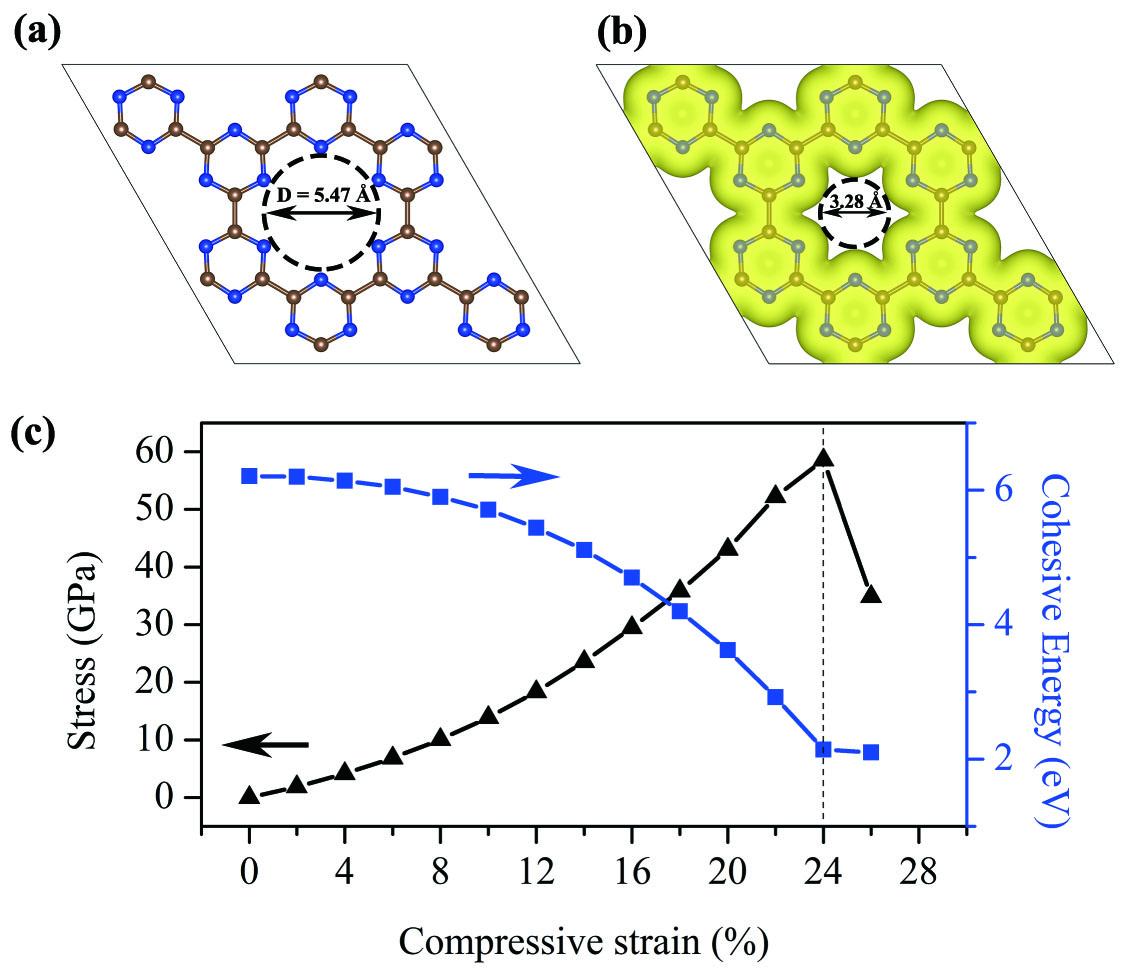}
   \caption{(a) Fully optimized $2\times2$ supercell of CN sheet. (b) Electron density isosurface of CN sheet with an isovalue of 0.015~e/\AA$^{\rm 3}$. The brown and blue balls represent the C and N atoms, respectively. (c) The stress (black triangles) and the cohesive energy (blue squares) as a function of strain. The long dashed lines between the symbols are guides to the eyes. The vertical short dashed line indicates the critical strain.}
\end{figure}

\begin{figure}[!tp]
  \centering
  \includegraphics[width=9cm]{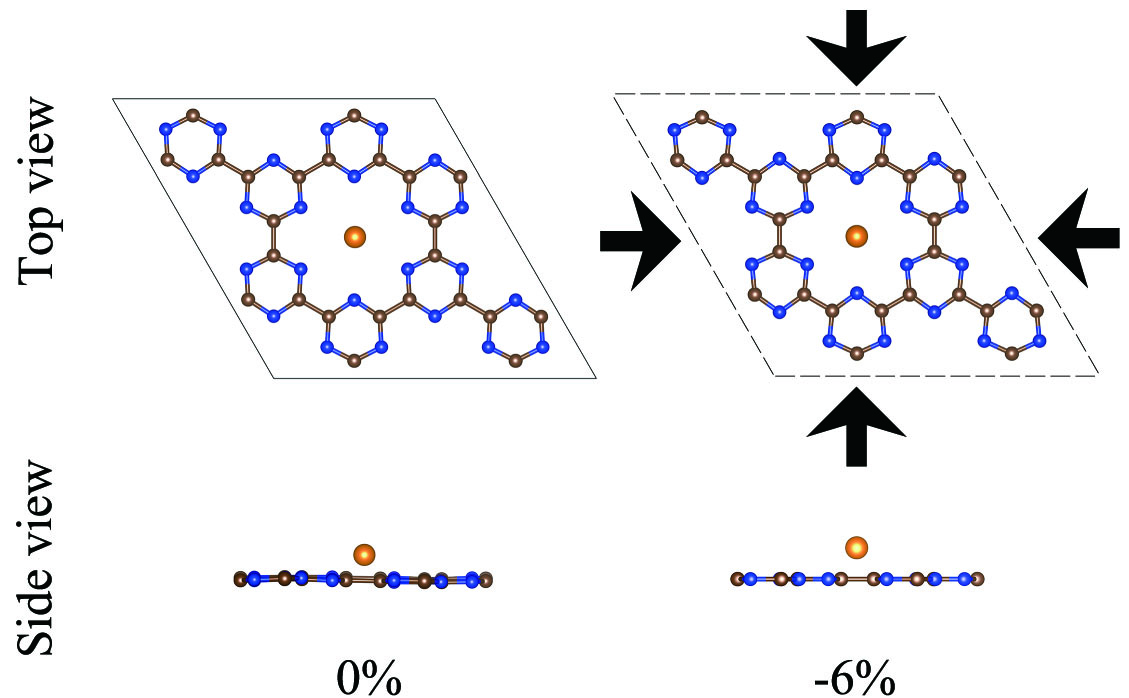}
  \caption{The geometrical structures of He adsorbed on CN sheet in the absence and presence of strain.
  The yellow, brown and blue balls represent He, C and N atoms, respectively.}
\end{figure}

\begin{figure}[!tp]
  \centering
  \includegraphics[width=9cm]{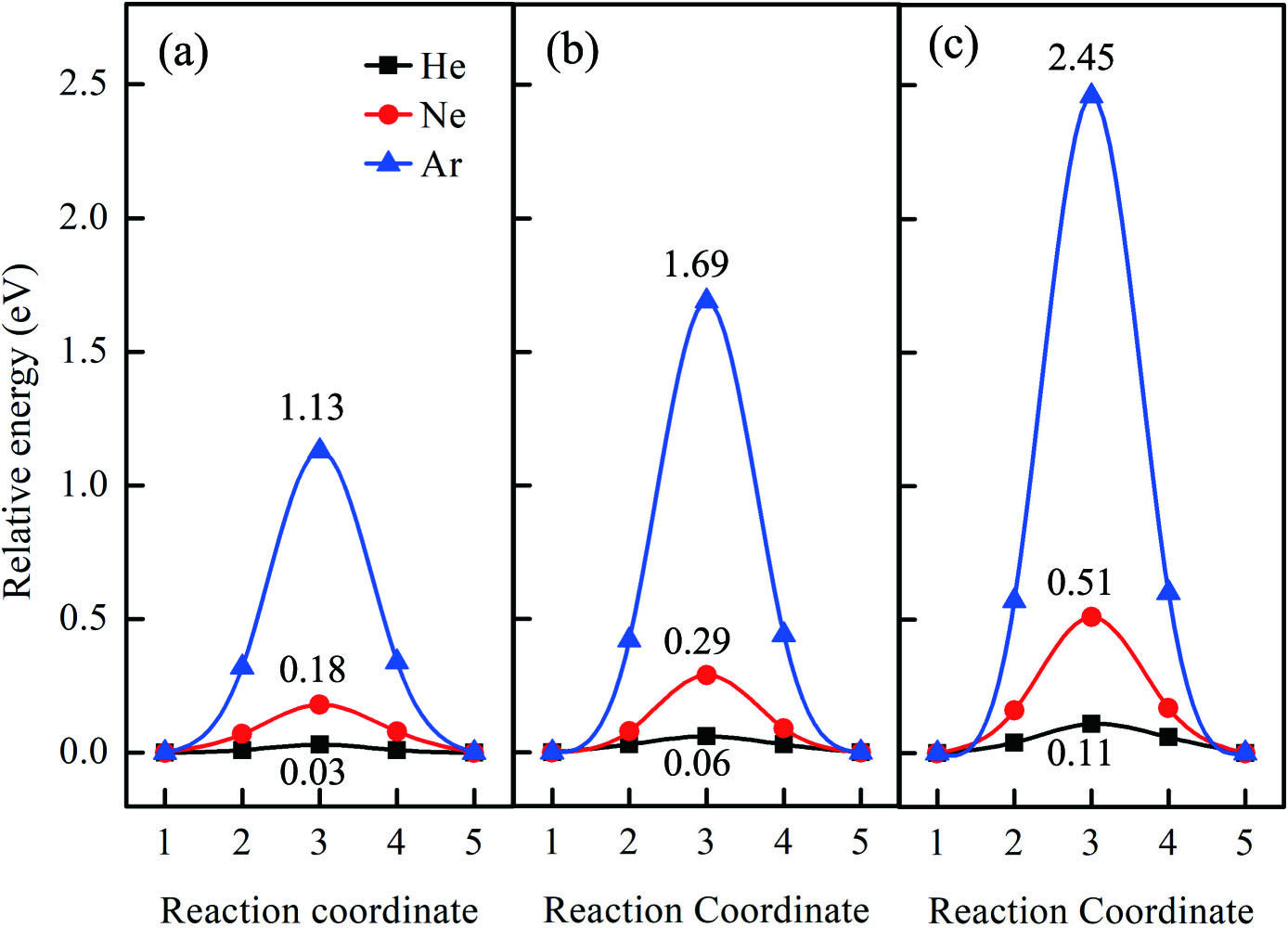}
  \caption{Energy profiles for He, Ne and Ar penetrating through CN membrane under (a) 0\%, (b) $-$3\% and (c) $-$6\% strain.}
\end{figure}

\begin{figure}
  \centering
  \includegraphics[width=9cm]{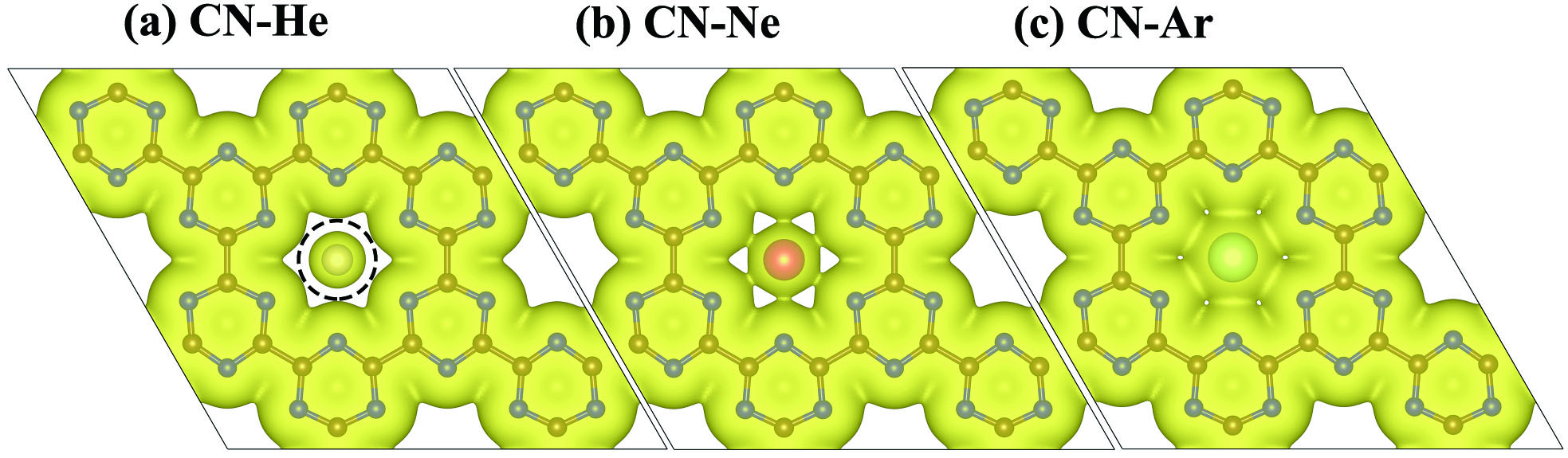}
  \caption{Electron density isosurface for (a) He, (b) Ne and (c) Ar at transition state under $-$6\% strain. The isovalue is 0.015~e/\AA$^{\rm 3}$.}
\end{figure}

\begin{figure}[!tp]
  \centering
  \includegraphics[width=9cm]{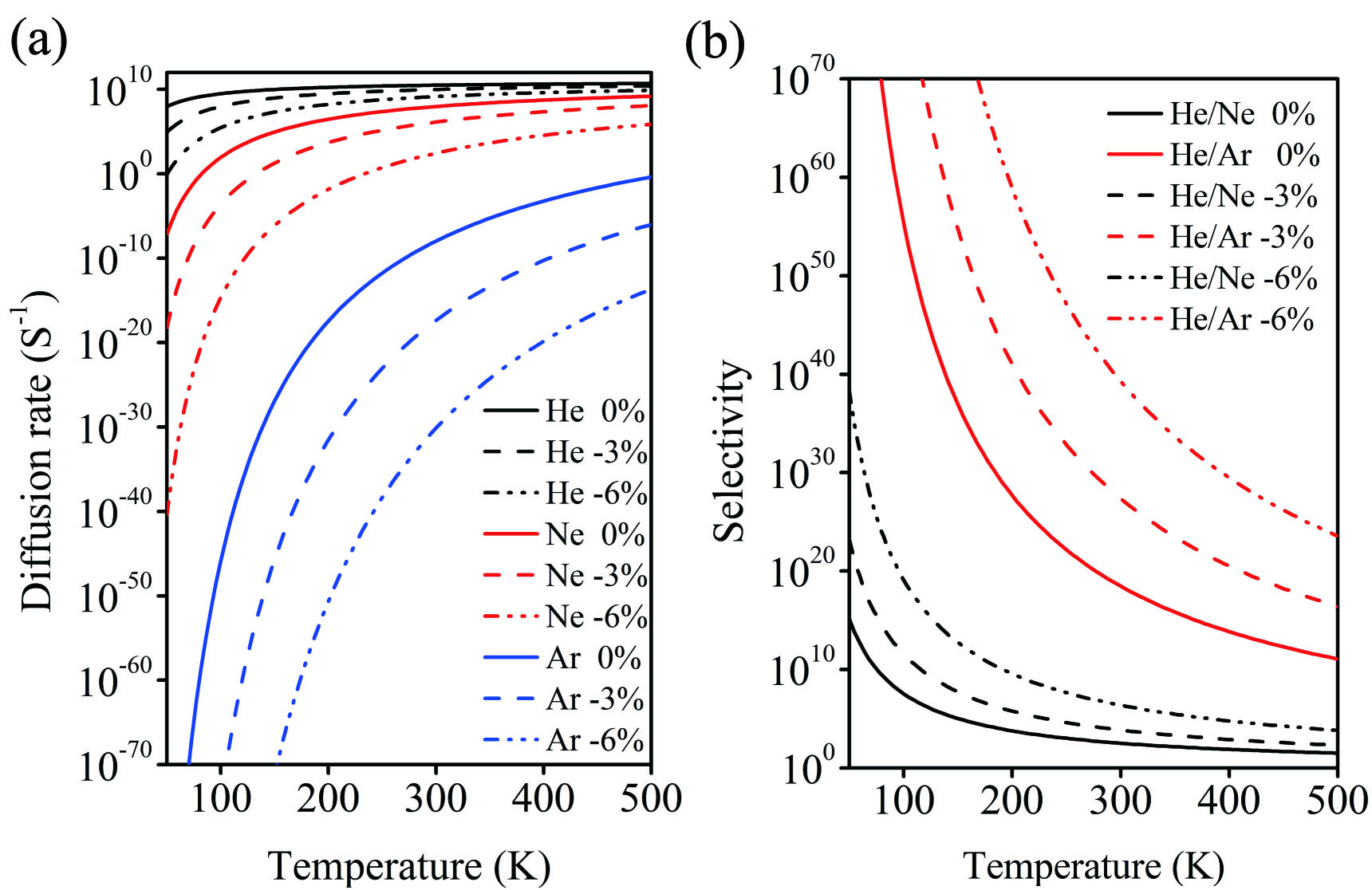}
  \caption{(a) Diffusion rate for the noble gas molecules, and (b) the selectivity of He relative to Ne and Ar molecules, as a function of temperature.}
\end{figure}

\begin{figure}
  \centering
  \includegraphics[width=6cm]{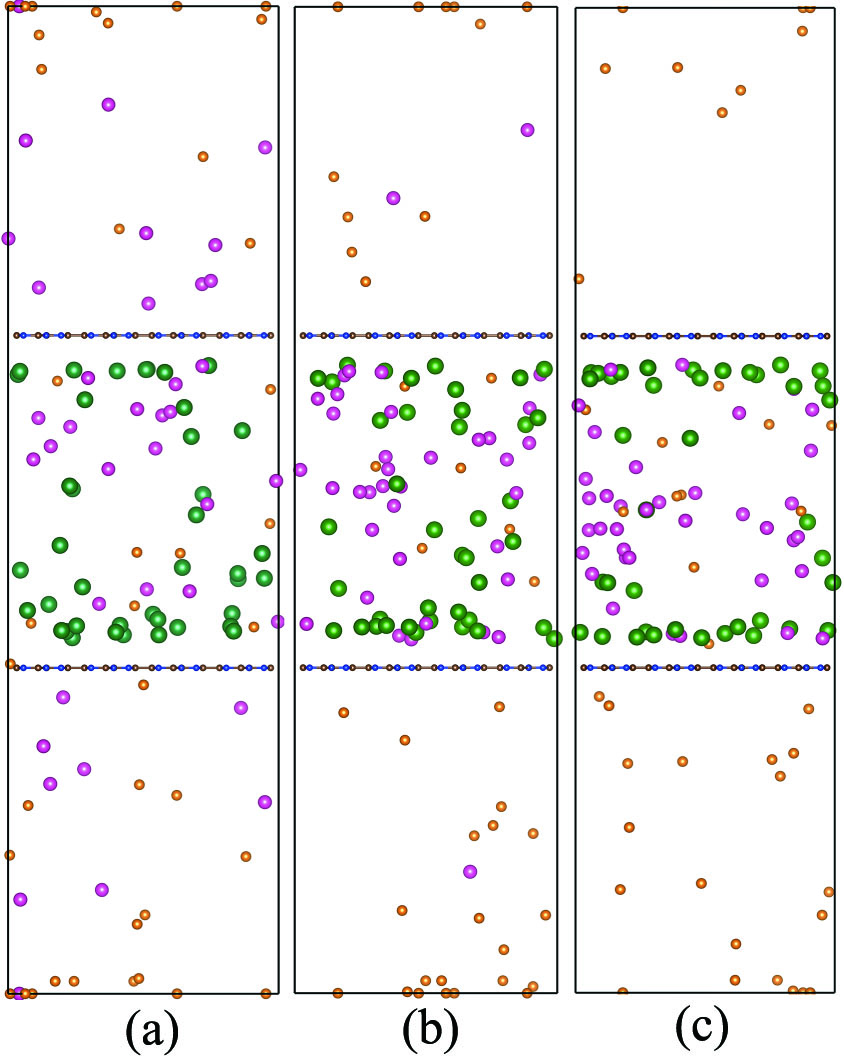}
  \caption{Final configuration of the mixed gases permeating through the CN membrane at room temperature under a strain of (a) 0\%, (b) $-$3\% and (c) $-$6\%, respectively. Color code: yellow, He; pink, Ne; green, Ar. }
\end{figure}

\begin{table}
\caption{\label{tabone} Kinetic diameter D$_{\rm 0}$~(\AA) of the gas molecules, the interaction energy $E_{\rm int}$ (meV) and the adsorption height H$_{\rm 0}$ (~\AA) of CN membrane with gas molecules.}
\begin{tabular}{p{1.6cm}p{1.6cm}p{1.6cm}p{1.6cm}p{1.6cm}}
\hline
 &D$_{\rm 0}$  & Strain  & $E_{\rm {int}}$ & H$_{\rm 0}$ \\
 \hline
 & &  \ \   0\% &   $-$9.5 & 1.05 \\
          He &  2.6       & $-$3\% &  $-$18.0 &  1.26  \\
           &           & $-$6\%   & $-$25.5  & 1.33  \\
\hline
  & &    \ \ 0\% &   $-$7.1  & 1.64 \\
         Ne    &    3.2     & $-$3\%  & $-$17.6  & 1.81  \\
             &      &   $-$6\% &  $-$24.8  & 1.92  \\
\hline
& &   \ \ 0\% &  $-$6.9 & 2.42 \\
         Ar   &     3.4     & $-$3\%   & $-$16.3  & 2.56  \\
             &        & $-$6\% &   $-$24.0  & 2.61  \\
\hline
\end{tabular}
\end{table}

\begin{table}
\caption{\label{tabone} The selectivity (S) of He toward Ne and Ar penetrating through CN membrane under $-$6\% strain at room temperature.
The results of previously studied membranes C$_{\rm 2}$N and silicene are also included for comparison.}
\begin{tabular}{p{2cm}p{2cm}p{2.8cm}p{2.8cm}}
\hline
  & CN  & C$_{\rm 2}$N (Ref.\cite{Zhu2015C2N}) &  Silicene (Ref.\cite{Hu2013Helium})  \\
 \hline
 S(He/Ne) & $5.17\times10^6$ & $3\times10^3$ & $2\times10^3$  \\
 S(He/Ar) & $1.89\times10^{39}$ & $4\times10^{18}$ & $1\times10^{18}$  \\
\hline
\end{tabular}
\end{table}

\begin{table}
\caption{\label{tabone} The He permeance (GPU) of CN monolayer under $-$6\% strain at 300~K, and the comparison with those of porous graphene (PG) and C$_{\rm 2}$N.
$\left [ 1~{\rm GPU} = 3.35\times10^{-10}~mol/(m^2 \cdot s \cdot Pa) \right ]$}
\begin{tabular}{p{2cm}p{2cm}p{2.8cm}p{2.8cm}}
\hline
            & CN  & C$_{\rm 2}$N (Ref.\cite{Zhu2015C2N}) &  PG (Ref.\cite{Brockway2013Noble}) \\
\hline
 Permeance & $1.94\times10^7$ & $1\times10^7$ & $7\times10^4$  \\
\hline
\end{tabular}
\end{table}

\end{document}